# Quantum mechanics of hyperbolic metamaterials: Modeling of quantum time and Everett's "universal wavefunction"

Igor I. Smolyaninov

*Department of Electrical and Computer Engineering, University of Maryland, College Park, MD 20742, USA*

*phone: 301-405-3255;   fax: 301-314-9281;   e-mail: smoly@umd.edu*

**Modern advances in transformation optics and electromagnetic metamaterials made possible experimental demonstrations of highly unusual curvilinear "optical spaces", such as various geometries necessary for electromagnetic cloaking. Recently we demonstrated that mapping light intensity in a hyperbolic metamaterial may also model the flow of time in an effective (2+1) dimensional Minkowski spacetime. Curving such an effective spacetime creates experimental model of a toy "big bang". Here we demonstrate that at low light levels this model may be used to emulate a fully covariant version of quantum mechanics in a (2+1) dimensional Minkowski spacetime. When quantum mechanical description is applied near the toy "big bang", the Everett's "universal wave function" formalism arises naturally, in which the wave function of the model "universe" appears to be a quantum superposition of mutually orthogonal "parallel universe" states.**

# 1. Introduction

The problem of time in quantum mechanics has a long history. In the standard non-relativistic quantum mechanics time and space variables are treated differently: time is treated as a parameter, not an operator, and the uncertainty principle for time and energy has a different character than the uncertainty principle for space and momentum. Relativistic extensions of standard quantum mechanics do not resolve these issues completely. The main difficulty is associated with the role of measurement in quantum mechanics, which is supposed to collapse the wave function "instantaneously". Quite obviously, the notion of instantaneous collapse appears to be difficult to accommodate in a relativistic theory. The "peaceful coexistence" of quantum mechanics and relativity is especially difficult to arrange when measurements performed on spatially separated two-particle entangled states are considered. Various interpretations of quantum mechanics do not agree on these issues. Moreover, they often predict different experimentally testable outcomes. A good recent overview of the notion of time in quantum mechanics can be found in ref.[1] and the references therein.

Electromagnetic metamaterials appear to provide us with an interesting model system, which may enable better understanding of the meaning of time in various physics situations. Recently we demonstrated that mapping of monochromatic extraordinary light distribution in a hyperbolic metamaterial along some spatial direction may model the classical "flow of time" [2,3]. Moreover, it seems to be possible to examine simultaneous appearance of "statistical" and "cosmological" arrows of time in an experimental scenario which emulates a big bang-like event in a specially designed hyperbolic metamaterial [3]. Here we examine optical properties of this model at low light levels, and demonstrate that it may be used to emulate a fully covariant version of quantum mechanics in a 2+1 dimensional Minkowski spacetime in which the "time" and "space" operators enter all equations in a completely symmetric fashion. An interesting feature of this model is that the rules of standard non-relativistic quantum mechanics allow us to build a simple and non-controversial model of Lorentz invariant quantum theory in an effective flat (2+1) Minkowski spacetime.

Hyperbolic metamaterials are typically fabricated as either layered metal-dielectric structures or arrays of metal wires in a dielectric host [2,3]. On the other hand, one of the best understood geometries for hyperbolic metamaterials relies on $In_{0.53}Ga_{0.47}As:Al_{0.48}In_{0.52}As$ semiconductor superlattices [4], making this topic of particular interest to semiconductor research community. Moreover, since the beginning of 80s quantum-optical analogies in various photonic and semiconductor settings were extensively discussed in non-relativistic case in the Schrödinger approximation, as described in recent extensive reviews [5-7]. Such fundamental for solid-state quantum physics effects as Bloch oscillations, Zenner tunneling, surface Tamm states, Anderson localization and so on were modeled in micro-structured materials and photonic lattices. In addition, relativistic case was also addressed in refs. [8-10]. Metamaterial modeling of relativistic fully covariant quantum mechanics builds on this extensive body of work.

**2. Modeling a fully covariant version of quantum mechanics in a (2+1) dimensional Minkowski spacetime**

Let us start by recalling the basic properties of the metamaterial model of time, which is described in detail in refs.[2,3]. We will consider a non-magnetic uniaxial anisotropic metamaterial, which has constant dielectric permittivities $\varepsilon_x = \varepsilon_y = \varepsilon_1 > 0$ and $\varepsilon_z = \varepsilon_2 < 0$ in some frequency range around $\omega = \omega_0$. Such a metamaterial is usually called "indefinite" or "hyperbolic". Any electromagnetic field propagating in this metamaterial may be expressed as a sum of the "ordinary" and "extraordinary" contributions depending on the vector $\vec{E}$ direction with respect to optical axis. $\vec{E}$ is perpendicular to the optical axis for the ordinary component of the electromagnetic field, while extraordinary photons have nonzero $\vec{E}$ component along the optical axis. Let us introduce a "scalar" extraordinary field as $\varphi = E_z$. Let us assume that the metamaterial is illuminated by high intensity coherent CW laser field at frequency $\omega_0$, and we study spatial distribution of the extraordinary field $\varphi_\omega$ at this frequency. Since temporal dispersion is rather large in hyperbolic metamaterials, we need to work in the frequency domain. Therefore, macroscopic Maxwell equations can be written as

$$\frac{\omega^2}{c^2}\vec{D}_\omega = \vec{\nabla}\times\vec{\nabla}\times\vec{E}_\omega \quad \text{and} \quad \vec{D}_\omega = \vec{\vec{\varepsilon}}_\omega \vec{E}_\omega \quad (1)$$

After simple transformations Eq.(1) results in the following equation for $\varphi_\omega$ field:

$$-\frac{1}{\varepsilon_1}\frac{\partial^2 \varphi_\omega}{\partial z^2} + \frac{1}{|\varepsilon_2|}\left(\frac{\partial^2 \varphi_\omega}{\partial x^2} + \frac{\partial^2 \varphi_\omega}{\partial y^2}\right) = \frac{\omega_0^2}{c^2}\varphi_\omega = \frac{m^{*2}c^2}{\hbar^2}\varphi_\omega, \quad (2)$$

where $m^* = \hbar\omega_0/c^2$ plays the role of effective mass. Equation (2) looks similar to the 3D Klein-Gordon equation describing a massive scalar $\varphi_\omega$ field. Spatial coordinate $z = \tau$ behaves as a "timelike" variable in this equation. Thus, it is clear that at large illumination levels eq.(2) describes propagation of light rays which behave as world lines of massive particles in a flat (2+1) dimensional Minkowski spacetime [3]. For example, if a dipole source (say a dye molecule) oscillating at frequency $\omega_0$ is placed inside the dielectric phase of the hyperbolic metamaterial, its radiation pattern looks like a light cone in Minkowski spacetime (see Fig.1).

Let us now reduce the illumination level (using e.g. a neutral density filter positioned in between the CW laser and the hyperbolic metamaterial sample) and examine the transition from ray optics to quantum mechanical propagation of photons inside the hyperbolic metamaterial. The transition from classical to quantum optics occurs when the number of photons N in any given mode is no longer large, so that the assumption of small fluctuations $N^{1/2} \ll N$ is no longer valid. Using the correspondence principle, the wave equation describing extraordinary photons propagating inside the hyperbolic metamaterial may be re-written as follows:

$$\left(-\frac{1}{\varepsilon_1}\frac{\partial^2}{\partial z^2}+\frac{1}{|\varepsilon_2|}\left(\frac{\partial^2}{\partial x^2}+\frac{\partial^2}{\partial y^2}\right)-\frac{m^{*2}c^2}{\hbar^2}\right)\varphi_\omega=0 \qquad (3)$$

where $\varphi_\omega$ is understood as a quantum mechanical photon wave function. In the "non-relativistic" limit in which the second term (the kinetic energy) in the parenthesis is much smaller than the effective rest energy $m^*c^2$, eq.(3) reduces to a standard Schrödinger equation:

$$i\frac{\hbar c}{\varepsilon_1}\frac{\partial}{\partial z}\varphi_\omega=\hat{H}\varphi_\omega=\pm\left(m^*c^2+\frac{\hbar^2}{|\varepsilon_2|2m^*}\left(\frac{\partial^2}{\partial x^2}+\frac{\partial^2}{\partial y^2}\right)\right)\varphi_\omega=\pm\left(m^*c^2+\frac{(\hat{p}_x^2+\hat{p}_y^2)}{|\varepsilon_2|2m^*}\right)\varphi_\omega$$

(4)

Note that if we would allow $\varepsilon_1$ and $\varepsilon_2$ to vary, an effective potential energy term would also appear in eq.(4). The $m^*c^2$ term is usually omitted in the non-relativistic quantum mechanics by re-defining zero energy. Moreover, since the experimental geometry is stationary (we use coherent CW laser illumination so nothing dependents on time), and the Maxwell equations do not change under $k_z \rightarrow -k_z$ transformation, we may choose $k_z$ to be positively defined. As a result, eq.(4) replicates standard non-relativistic quantum mechanics in the (2+1) Minkowski spacetime in which $z=\tau$ coordinate plays the role of effective time. In a more general "relativistic" situation where the effective kinetic energy is no longer much smaller than the rest energy $m^*c^2$, the "relativistic" eq.(3) must be used. Note that the "temporal" coordinate $z=\tau$ and the spatial coordinates $x$ and $y$ enter "relativistic" wave equation (3) in a completely symmetric fashion. Therefore, within the scope of our model the uncertainty principle between $z=\tau$ and $k_z=E$ acquires exactly the same meaning as the uncertainty principle between $x$ and $k_x=p$. On the other hand, definition of $k_z$ as always positive in the "non-relativistic limit" of our model would be clearly responsible for the re-introduction of all the typical difficulties associated with the notion of time in quantum mechanics [1].

The only missing ingredient in our model of (2+1) dimensional Lorentz invariant quantum theory remains a notion of classical particle detector. Ideally, such a detector must be an "internal" one. It would need to be made of some combination of our model photon "particles", and therefore, it would need to rely on nonlinear interaction of extraordinary photons. Potentially, this may be achieved in a hyperbolic metamaterial which exhibits strong optical nonlinearity and form spatial solitons [11]. Solitons may scatter other photons due to changes in local dielectric permittivity, leading to "internal measurement" of the photon spacetime location. On the other hand, in the absence of nonlinearity the role of classical detectors may be played by structural defects of the metamaterial, which would scatter extraordinary photons at a given $(z_0=\tau_0, x_0, y_0)$ "spacetime location". Once again, these scattering events may be considered as measurements of the particle spacetime location. For example, in the 2D plasmonic hyperbolic metamaterial model described in [3] a near-field optical probe may be used as such a detector. The probe may be positioned at any desired location above the 2D metamaterial leading. It would scatter photons at this location. Construction of a "particle detector" completes our model of Lorentz invariant quantum theory in (2+1) dimensions, which is non-local by design. This model may be built in an actual experiment and compared with our own world.

Even though our model is limited by the fact that it is populated by scalar particles only, this does not prevent us from consideration of the most interesting implications of the quantum time concept, such as the temporal two slit experiments [1] and the issues of post-selection [12]. By design, the results of temporal two slit experiments in our model would replicate the results of quantum time model described in ref.[1], since in both models the particle wave function is extended both along spatial and temporal directions (Fig.2). Additional dispersion predicted by the quantum time model [1] should be easy to observe in the model experiments with hyperbolic metamaterials. As described in ref.[1], in standard quantum theory, a particle going through a gate in time is clipped by the gate, reducing its dispersion in time. On the other hand, in temporal quantization, in addition to clipping, the particle is diffracted by the gate as well, increasing its dispersion in time relative to the standard quantum theory result. Diffraction of the photon by a "temporal gate" [1] in the $z=\tau$ direction would indeed be easy to observe in model experiments with hyperbolic metamaterials, while widened interference peaks would be also easy to observe in an analogue of the temporal double slit experiment [1]. It is also obvious that within the scope of our model, measurements performed in the "past" and in the "future" must be treated on absolutely equal footing as long as the experimental arrangement is symmetric with respect to $k_z \rightarrow -k_z$ transformation, and the metamaterial spacetime is flat, so that our model quantum mechanics is perfectly "time symmetric" and both pre-selection and post selection of the quantum states is allowed.

## 3. Modeling the Everett's universal wave function formalism

On the other hand, such a symmetric experimental arrangement may be deformed, so that a preferred "cosmological arrow of time" may be created [3]. If $\varepsilon_1$ and $\varepsilon_2$ are allowed to vary, the Klein-Gordon equation for a massive particle in a gravitational field

$$\frac{1}{\sqrt{-g}}\frac{\partial}{\partial x^i}\left(g^{ik}\sqrt{-g}\frac{\partial \varphi}{\partial x^k}\right) = \frac{m^2 c^2}{\hbar^2}\varphi \qquad (5)$$

may be emulated in a hyperbolic metamaterial. For example, let us consider an experimental situation, in which we allow slow adiabatic variation of $\varepsilon_2$ as a function of $z$, while $\varepsilon_1$ is kept constant. According to eq.(2) this situation corresponds to "cosmological expansion" of the (2+1) dimensional universe as a function of "timelike" $z=\tau$ variable. Simple experimental models of standard and inflationary big bang cosmologies based on this idea have been demonstrated in refs.[3,13] using plasmonic hyperbolic metamaterials. Hyperbolic metamaterial geometry in these demonstrations exhibited circular symmetry, and the radial coordinate $r$ played the role of "timelike" variable, so that the central point $r=\tau=0$ may be considered as a toy "big bang" (Fig.3). Field propagation far from the center of the concentric metamaterial structure may be described locally using rectangular coordinates so that radial coordinate $r$ corresponds to $z$, and angular coordinates correspond to $x$ and $y$ directions. Choosing $\varepsilon_1$=const>0 and $\varepsilon_2 \sim -e^{Hz}$ (where $H$ is the effective "Hubble constant") reproduces eq.(5) for massive particles if we introduce new wave function $\psi$ as $\psi=(-\varepsilon_2)^{1/2}\varphi_\omega$. Indeed, such a substitution leads to the following wave equation for $\psi$:

$$-\frac{\partial^2\psi}{\partial z^2}+\frac{\varepsilon_1}{e^{Hz}}\left(\frac{\partial^2\psi}{\partial x^2}+\frac{\partial^2\psi}{\partial y^2}\right)-H\left(\frac{\partial\psi}{\partial z}\right)=\left(\frac{\varepsilon_1\omega_0^2}{c^2}+\frac{H^2}{4}\right)\psi \qquad (6)$$

which differs from eq.(5) only by the scaling factor $\varepsilon_1$ in the xy-direction, and is valid at any large $r=z$ [13]. At low illumination level this experimental geometry may be used to emulate quantum mechanics on the inflationary de Sitter spacetime background. Using the correspondence principle, eq.(6) may be re-written as

$$\left(-\frac{\partial^2}{\partial z^2}+\frac{\varepsilon_1}{e^{Hz}}\left(\frac{\partial^2}{\partial x^2}+\frac{\partial^2}{\partial y^2}\right)-H\frac{\partial}{\partial z}-\left(\frac{\varepsilon_1\omega_0^2}{c^2}+\frac{H^2}{4}\right)\right)\psi=0 \qquad (7)$$

where the wave function $\psi$ may be interpreted as the wave function of the toy metamaterial universe. It plays the same role as the Everett's universal wavefunction [14]. By construction such a model breaks symmetry with respect to $k_z \rightarrow -k_z$ transformation, which is also obvious from the appearance of $H\partial/\partial z$ term in eq.(7). Breaking this symmetry defines the "future time direction" in our metamaterial model.

Fabrication of the plasmonic hyperbolic metamaterial shown in Fig.3(b,c) requires only very simple and common lithographic techniques. Let us consider a surface plasmon (SP) wave which propagates over a flat metal-dielectric interface. In our experiments gold has been chosen as a good plasmonic metal, while thin layers of PMMA have been used as a dielectric. If the metal film is thick, the SP wave vector is defined by expression

$$k_p = \frac{\omega}{c}\left(\frac{\varepsilon_d \varepsilon_m}{\varepsilon_d + \varepsilon_m}\right)^{1/2} \qquad (8)$$

where $\varepsilon_m(\omega)$ and $\varepsilon_d(\omega)$ are the frequency-dependent dielectric constants of the metal and dielectric, respectively [15]. Let us introduce an effective 2D dielectric constant $\varepsilon_{2D}$ so that $k_p = \varepsilon_{2D}^{1/2}\omega/c$, and thus

$$\varepsilon_{2D} = \left(\frac{\varepsilon_d \varepsilon_m}{\varepsilon_d + \varepsilon_m}\right) \qquad (9)$$

Now it is easy to see that depending on the frequency, SPs perceive the dielectric material bounding the metal surface in drastically different ways. At low frequencies $\varepsilon_{2D} \approx \varepsilon_d$. Therefore, plasmons perceive a PMMA stripe as dielectric. On the other hand, at high enough frequencies around $\lambda_0 \sim 500$ nm, $\varepsilon_{2D}$ changes sign and becomes negative since $\varepsilon_d(\omega) > -\varepsilon_m(\omega)$. As a result, around $\lambda_0 \sim 500$ nm plasmons perceive PMMA stripes on gold as if they are "metallic layers", while gold/vacuum portions of the interface are perceived as "dielectric layers". Thus, at these frequencies plasmons perceive a PMMA stripe pattern from Fig.3(b) as a layered hyperbolic metamaterial having opposite signs of $\varepsilon$ along the radial and azimuthal directions. Such a metamaterial geometry indeed breaks symmetry with respect to $k_r \rightarrow -k_r$ transformation.

Note that rigorous theoretical description of the PMMA-based plasmonic metamaterials developed in ref. [16] produces similar answer.

Eq.(7) describes evolution of the universal wavefunction $\psi$ over the resulting (2+1) dimensional expanding spacetime. It is natural to express $\psi$ as a superposition of spacetime location eigenstates $\xi_m(r=\tau, x, y)$ shown schematically in Fig.3(c):

$$\psi = \sum c_m \xi_m(r=\tau, x, y) \tag{8}$$

The Gaussian width of the location eigenstates is chosen to be equal to the width of an individual plasmonic ray or "world line" in Fig.3(c). When the CW laser power is attenuated $\sum c_m^2 = N$, where $N$ is the total (small) number of extraordinary photons launched into the metamaterial. Unless we perform measurements of the photon space-time locations using e.g. a near-field optical probe as described above, any superposition of the location eigenstates described by eq.(8), which satisfies $\sum c_m^2 = N$ condition exists simultaneously in the quantum mechanical sense. Therefore, within the scope of our model such location eigenstate superpositions may be considered as co-existing "parallel universes". By taking into account results of all the spacetime location measurements performed on the system, these "parallel universes" may be arranged as sets of "consistent histories". Thus, all the features of the Everett's universal wave function formalism arise naturally within the scope of our model, and by construction our model is free from any inconsistency or paradox.

**4. Conclusion**

In conclusion, we have demonstrated that propagation of low intensity light through a hyperbolic metamaterial may be used to emulate a fully covariant version of quantum mechanics in a (2+1) dimensional effective Minkowski spacetime. When quantum mechanical description is applied near the toy "big bang", the Everett's "universal wave function" formalism arises naturally, in which the wave function of the model

"universe" appears to be a quantum superposition of mutually orthogonal "parallel universe" states.

**Figure Captions**

**Figure 1.** Radiation pattern of a dipole source placed inside a hyperbolic metamaterial looks like a light cone in a 2+1 dimensional Minkowski spacetime in which spatial z coordinate plays the role of a "timelike" coordinate.

The calculations have been performed using COMSOL Multiphysics 4.2.

**Figure 2.** Wave functions of photons in a hyperbolic metamaterial depend on both "temporal" $z=\tau$ and spatial $x$ and $y$ coordinates. This corresponds to the "quantum time" model of quantum mechanics described in ref.[1].

**Figure 3**. Quantum mechanical description of the "metamaterial spacetime": (a) Schematic view of world lines behavior near the big bang. (b) AFM image of the plasmonic hyperbolic metamaterial based on PMMA stripes on gold. The defect used to launch plasmons into the structure near the "big bang" location is shown by an arrow. (c) Plasmonic rays or "world lines" increase their spatial separation as a function of "timelike" radial coordinate. The point (or moment) $r=\tau=0$ corresponds to a toy "big bang". The spacetime location eigenstates are shown by circles. For the sake of clarity, light scattering by the edges of the PMMA pattern is partially blocked. (d) Proposed geometry of the "quantum time" experiments. Neutral density (ND) filter attenuates intensity of light coupled into the toy (2+1) dimensional spacetime. Unless we perform measurements of the photon space-time locations using e.g. a near-field optical probe(s) any superposition of the location eigenstates described by eq.(8) "exists" simultaneously in the quantum mechanical sense. Therefore, such superpositions of location eigenstates may be considered as "parallel universes" within the scope of our model.

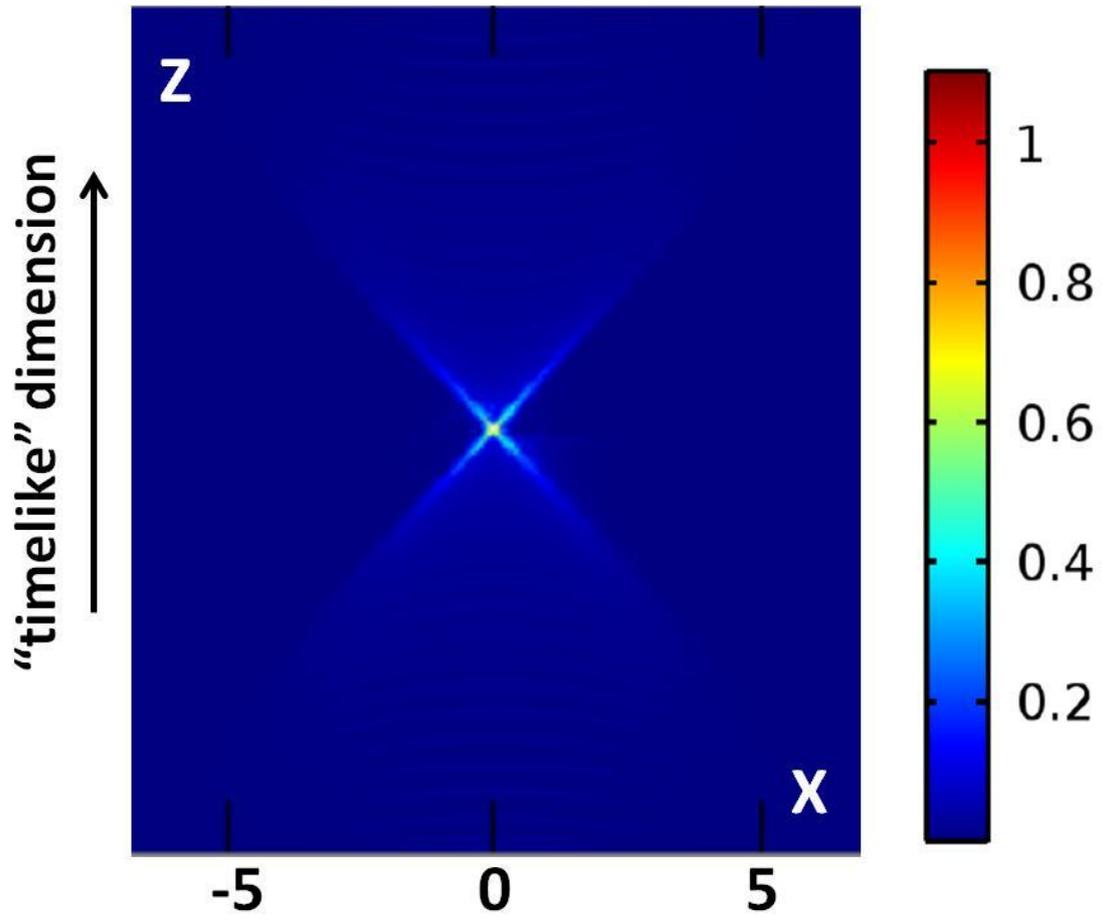

Fig.1

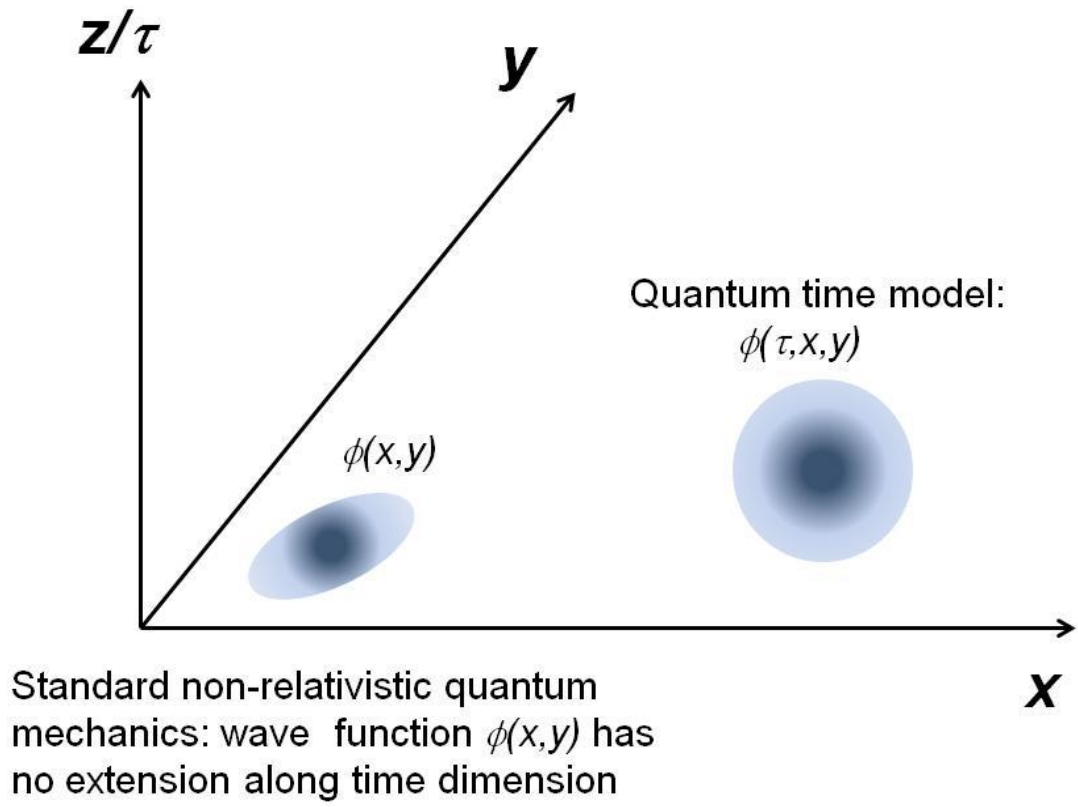

Fig.2

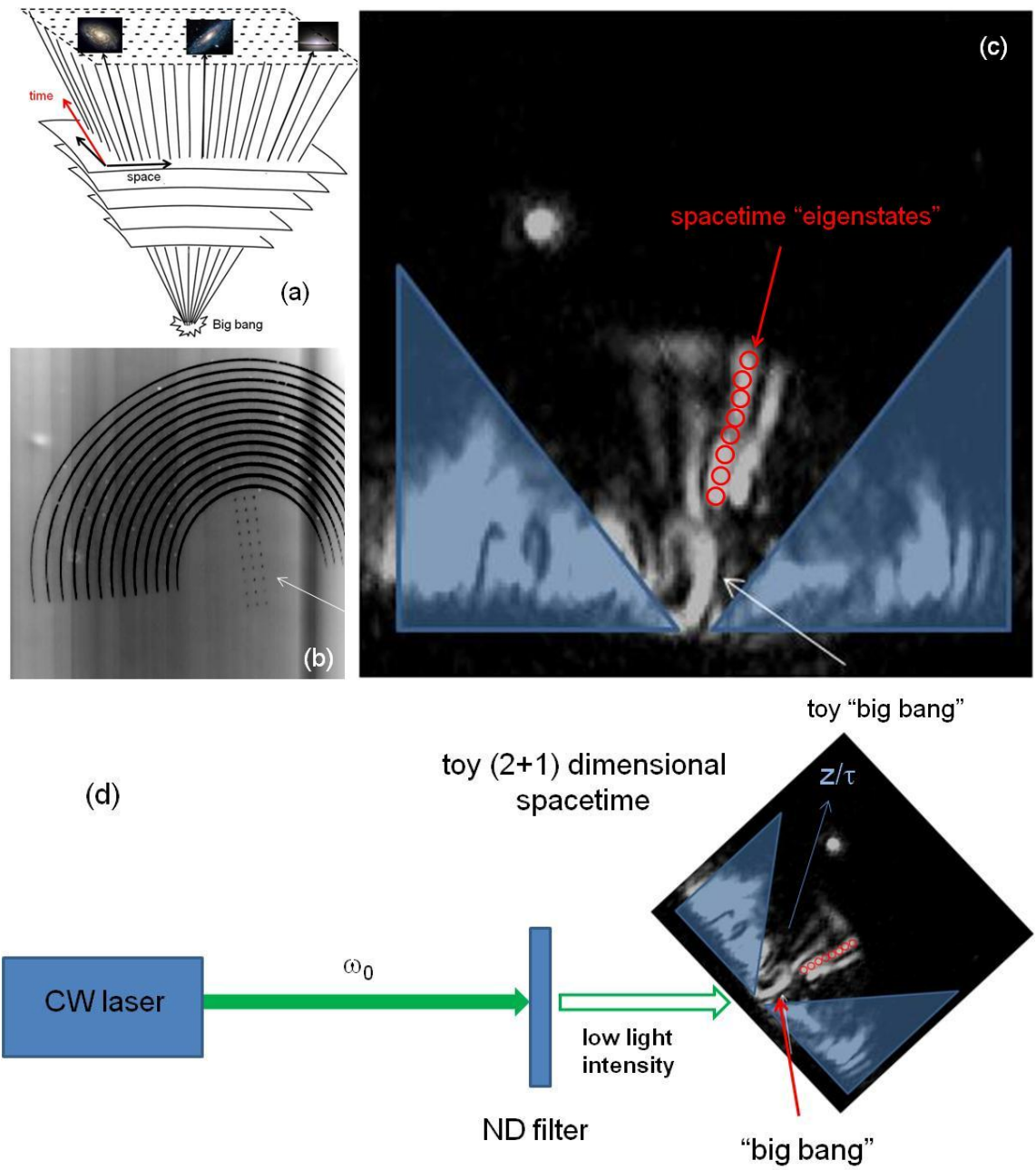

Fig.3